\documentclass[12pt]{iopart}
\usepackage{graphicx}
\usepackage{dcolumn}
\usepackage{bm}
\usepackage{siunitx}
\usepackage{multirow}
\usepackage{tikz}
\usepackage{hyperref}
\usepackage{caption}
\usepackage{subfig}      
\usepackage{threeparttable}
\usepackage{booktabs} 
\usepackage{cite}

\definecolor{a_blue}{HTML}{02BDD6}
\definecolor{a_purple}{HTML}{b793db}
\definecolor{a_grey}{HTML}{767690}
\definecolor{a_yellow}{HTML}{999900}

\begin{document}

\title{Thermodynamic stability and vibrational properties of multi-alkali antimonides}

\author{Julia Santana-Andreo$^{1,2}$, Holger-Dietrich Sa{\ss}nick$^1$, and Caterina Cocchi$^{1,2,3}$}
\address{$^1$ Carl von Ossietzky Universit\"at Oldenburg, Institute of Physics, 26129 Oldenburg, Germany}
\address{$^2$ Humboldt-Universit\"at zu Berlin, Physics Department and IRIS Adlershof, 12489 Berlin, Germany}
\address{$^3$ Carl von Ossietzky Universit\"at Oldenburg, Center for Nanoscale Dynamics (CeNaD), 26129 Oldenburg, Germany}
\ead{julia.santana.andreo@uni-oldenburg.de, caterina.cocchi@uni-oldenburg.de}
\vspace{10pt}

\begin{abstract}
Modern advances in generating ultrabright electron beams have unlocked unprecedented experimental advances based on synchrotron radiation. Current challenges lie in improving the quality of electron sources with novel photocathode materials such as alkali-based semiconductors. To unleash their potential, a detailed characterization and prediction of their fundamental properties is essential. In this work, we employ density functional theory combined with machine learning techniques to probe the thermodynamic stability of various alkali antimonide crystals, emphasizing the role of the approximations taken for the exchange-correlation potential. Our results reveal that the SCAN functional offers an optimal trade-off between accuracy and computational costs to describe the vibrational properties of these materials. Furthermore, it is found that systems with a higher concentration of Cs atoms exhibit enhanced anharmonicities, which are accurately predicted and characterized with the employed methodology.
\end{abstract}

%
%
%
%
%

\newpage
\section{Introduction}
Multi-alkali antimonides are semiconductors used as electron sources in particle accelerators~\cite{musu18nimprsasdae,scha23jmcc,moha23micromachines}. Despite the persisting challenges in producing high-quality samples~\cite{mamu16jvsta,gaow17aplmater,schm18prab}, the research on these materials is relentless. In terms of preparation, various recipes have been proposed based on co-evaporation~\cite{cult15prab,ding17jap,feng17jap} or sequential growth~\cite{schu13aplmater,ruiz14aplmater}. The analysis of the produced photocathodes has focused both on their \textit{operando} performance in accelerators~\cite{cult11apl,baza11apl,vecc11apl,dai20electronics,wang21sr} as well as on their physical properties~\cite{cocc19sr,panu21nimpra,ruse22prl}.
These experimental efforts have been recently complemented by several theoretical studies aiming at an in-depth understanding of the fundamental characteristics of the materials~\cite{kala10jpcs,guo14mre,murt16bms,cocc18jpcm,cocc20pssr-rrl,amad21jpcm,cocc21m,schi22prm,wu23jmca} and their photoemission performance~\cite{dimi17jap,jens18jap,anto20prb,nang21prb,saha23jap}. 

Computational research on photocathodes has been boosted by the advent of high-throughput screening methods which have enabled the exploration of large configurational spaces and the identification of optimal compounds for this specific application~\cite{capu16jem,cocc21m,sass22jcp}. So far, corresponding efforts have been focused mainly on determining relative stability and energy levels without the inclusion of thermodynamic effects. This choice, however, considerably limits the predictive power of such calculations. Notable exceptions are given by some recent first-principle works on the thermoelectric and vibrational properties of multi-alkali antimonides~\cite{yue22prb,shar23aaem} even beyond the harmonic approximation~\cite{zhon21ijer,yuan22jmcc}. Yet, the validity of the chosen approximations on a variety of structures and compositions needs to be carefully checked.

In this work, we present a first-principle study on the thermodynamic and vibrational properties of binary and ternary alkali antimonide crystals. After assessing the performance of three popular exchange-correlation functionals of density-functional theory in terms of accuracy and computational costs, we analyze the phonon band structures within the harmonic approximation and beyond in a machine-learning-based computational framework. We find that all considered materials are dynamically stable at room temperature except for Cs$_2$KSb and that compounds with a larger concentration of Cs atoms exhibit enhanced anharmonic effects. Our results contribute to the characterization and understanding of multi-alkali antimonides and offer valuable information to predict their photoemission yield.

\section{Computational methods}
The results presented in this work are obtained by solving the Kohn-Sham (KS) equations~\cite{kohn65pr} of density functional theory (DFT)~\cite{hohe64pr}: 
\begin{equation}
\left[ -\frac{1}{2}\nabla^2 + v_s(\mathbf{r}) \right] \psi_i(\mathbf{r}) = \varepsilon_i^{\text{KS}} \psi_i(\mathbf{r}).
\label{eq:KS}
\end{equation}
In Eq.~\ref{eq:KS}, formulated in atomic units, $\varepsilon_i^{\text{KS}}$ are the energy eigenvalues, $\psi_i(\mathbf{r})$ the wave functions, and $v_s(\mathbf{r})$ the effective KS potential including the contributions from the external, the Hartree, and the exchange-correlation (xc) potentials:
\begin{equation}
v_s(\mathbf{r}) = v_{\text{ext}}(\mathbf{r}) + v_{H}(\mathbf{r}) + v_{xc}(\mathbf{r}).
\end{equation}
In contrast to $v_{\text{ext}}$ and $v_{H}(\mathbf{r})$, the exact form of $v_{xc}$ is unknown.
Herein, we apply for this term the generalized gradient approximation (GGA) in the Perdew-Burke-Ernzerhof (PBE) parameterization~\cite{perd96prl}, the SCAN functional~\cite{sun15prl} based on the meta-GGA, which has proven good performance in various material classes~\cite{zhan18njp, chen17pnas,sassn21es}, and the range-separated hybrid functional HSE06~\cite{heyd03jcp,heyd06jcp}.
 
All DFT calculations are performed with the VASP software suite~\cite{kres93prb, kres96cms,kres96prb, kres99prb} implementing the projector-augmented wave method.
The investigated structures are fully optimized in terms of lattice vectors and atomic positions with a force threshold of \(10^{-7} \, \si{\electronvolt\per\angstrom}\). In the self-consistent field calculations, an energy threshold of \(10^{-9} \, \si{\electronvolt}\) and an energy cutoff of 500~\si{\electronvolt} are adopted. In line with earlier works~\cite{zhon21ijer,amad21jpcm}, the k-grids chosen to sample the Brillouin zones of all considered systems include $12\times 12\times 12$ points except for hexagonal NaK$_\mathrm{2}$Sb where a $8\times 8\times 4$ k-mesh is taken. For the calculations with the HSE06 functional, which are considerably more demanding than those with PBE and SCAN, the number of k-points is reduced to $2\times 2\times 2$. We checked that the accuracy of the lattice parameters is not affected by this choice. 
The calculation of the non-analytical component of the dynamical matrix requires the determination of Born effective charges and of the dielectric tensor from DFT. An extra supporting grid is incorporated to evaluate augmentation charges to minimize the impact of noise on the forces.

Interatomic force constants (IFC) are computed using the \texttt{hiPhive} package~\cite{erik19ats}, which integrates machine learning regression techniques with forces calculated from random atomic distortions within supercells~\cite{plat22cm,plat23jmca,bror22atas}.
Distorted $2\times 2\times 2$ supercells 
are generated by a Monte Carlo algorithm that penalizes displacements producing too short interatomic distances~\cite{fran20ncm}. We checked that 10 supercells are sufficient to obtain a root mean square deviation for the forces below 18~m\si{\electronvolt\per\angstrom}. The mean displacement amplitude for
each configuration is 0.25 \si{\angstrom}. To compute the Hellmann-Feynman forces, which are essential for obtaining anharmonic IFCs, DFT calculations are carried out in supercells with a $6\times 6\times 6$ k-grid for cubic systems and a $4\times 4\times 2$ k-mesh for the hexagonal one; with HSE06, such calculations are performed at $\Gamma$ only. 
Multilinear regression is applied to fit forces derived from DFT, using the recursive feature elimination algorithm~\cite{fran20ncm}. We evaluate the convergence of the cutoff distances in the force constant potential (FCP) model by examining variations in force errors and phonon frequencies. Cutoff distances of 9.2, 9.1, 6.4, 4.5, and 4.5~\si{\angstrom} for the second, third, fourth, fifth, and sixth order, respectively, are necessary to achieve the desired convergence for Cs$_\mathrm{3}$Sb with PBE. These cutoff distances are subsequently extrapolated based on coordination shells for calculations with the other xc functionals on the remaining binary systems. 
A wrapper code for \texttt{hiPhive} facilitates automation of distorted supercell creation, force calculation using VASP, and the construction of ML-based FCPs~\cite{hiPhiveWrapper2021}.
The evaluation of temperature-dependent anharmonic phonon frequencies is performed by integrating self-consistent phonon (SCPH) calculations combined with the second-order to sixth-order IFCs~\cite{esfa08prb,erre14prb}.
In the study of ternary systems, only second-order IFCs are required to accurately compute harmonic phonon dispersion. 

\section{Results}
\subsection{Structural properties}

\begin{figure*}
\includegraphics[width=\textwidth]{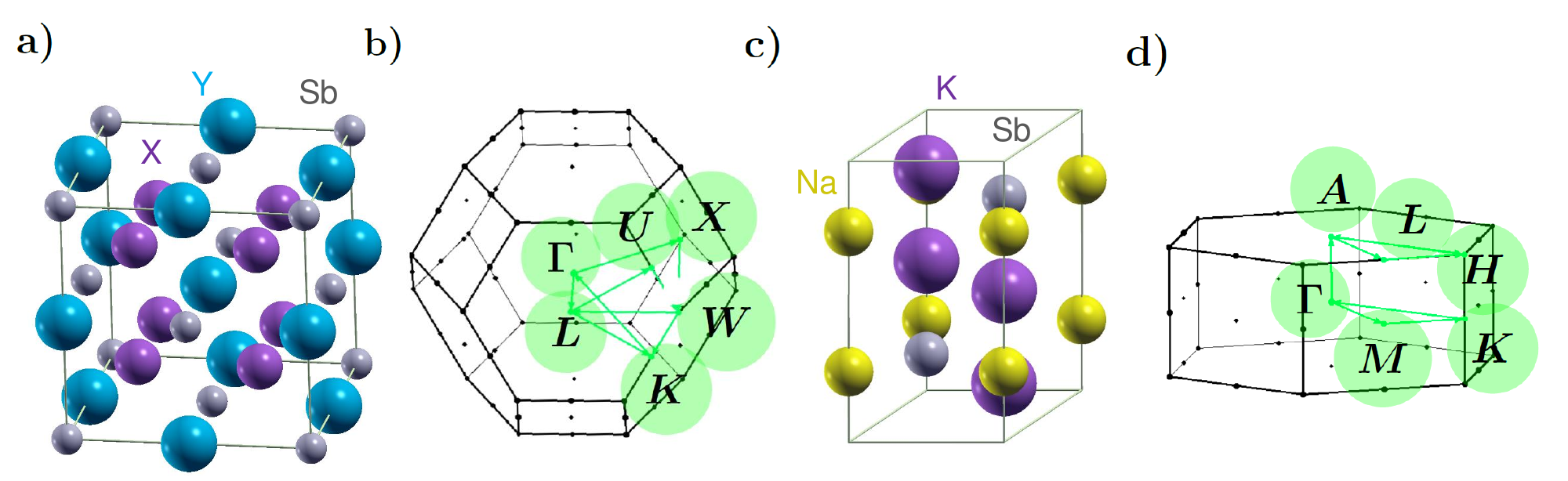} 
\caption{\label{fig:unitcell} a) Ball-and-stick representation of the conventional unit cell of multi-alkali antimonides with chemical formula X$_\mathrm{2}$YSb (X = Cs, K, Na; Y = Cs, K, Na) and b) corresponding Brillouin zone with the high-symmetry points and the path connecting them highlighted in green. c) Ball-and-stick representation of the conventional unit cell of NaK$_\mathrm{2}$Sb and d) its Brillouin zone with the high-symmetry points and the path connecting them highlighted in green. Graphics created with \texttt{XCrysDen} \cite{koka99jmgm}.}
\end{figure*}

In this study on multi-alkali antimonides, we consider three binary systems, Cs$_3$Sb, K$_3$Sb, and Na$_3$Sb, and four ternaries: Cs$_2$KSb, CsK$_2$Sb, Na$_2$Kb, and NaK$_2$Sb. 
All materials are modeled in their face-centered cubic (FCC) phase (space group $\textit{Fm3m}$) except for NaK$_\mathrm{2}$Sb which is hexagonal (space group $P6_3/mmc$).
The FCC crystals are simulated in their conventional unit cell including 16 atoms (see Figure~\ref{fig:unitcell}a with the corresponding Brillouin zone in Figure~\ref{fig:unitcell}b) where Sb atoms are located at the Wyckoff position 4a and the alkali elements at positions 4b and 8c.
In the ternary crystals, the sites with coordinates 4b and 8c are occupied by the second alkali species, as illustrated in Figure~\ref{fig:unitcell}a.
In the binary crystals, the alkali species on these sites are inequivalent with respect to the others~\cite{cocc20pssr-rrl}.
The hexagonal crystal structure of NaK$_\mathrm{2}$Sb is described in a conventional unit cell with 8 atoms (see Figure~ \ref{fig:unitcell}c and Brillouin zone in Figure~ \ref{fig:unitcell}d) with Na atoms at 2b positions, K atoms at 4f and Sb atoms at 4f in the Wyckoff notation.

\begin{table}[h]
\centering
\begin{threeparttable}
\caption{Lattice parameter (\(a\)) and unit-cell volume per atom (\(\Omega\)) of Cs$_\mathrm{3}$Sb, K$_\mathrm{3}$Sb, and Na$_\mathrm{3}$Sb, calculated using DFT with PBE, HSE06, and SCAN functionals and compared to available experimental references.}
\label{tbl:1}
\begin{tabular*}{0.9\textwidth}{@{\extracolsep{\fill}} lcccc}
\toprule
& PBE & HSE06 & SCAN & exp. \\
\midrule
\textbf{Cs$_\mathrm{3}$Sb} \\
\midrule
\(a\) [\si{\angstrom}] & 9.348 & 9.253 & 9.171 & 9.14-9.19\tnote{a} \\
\(\Omega\) [\si{\cubic\angstrom}/atom] & 50.81 & 49.51 & 48.08 & 47.72-48.51\tnote{a} \\
\midrule
\textbf{K$_\mathrm{3}$Sb} \\
\midrule
\(a\) [\si{\angstrom}] & 8.587 & 8.495 & 8.465 & 8.49\tnote{b} \\
\(\Omega\) [\si{\cubic\angstrom}/atom] & 39.57 & 38.32 & 37.91 & 38.25\tnote{b} \\
\midrule
\textbf{Na$_\mathrm{3}$Sb} \\
\midrule
\(a\) [\si{\angstrom}] & 7.308 & 7.278 & 7.447 & 7.40-7.46\tnote{c} \\
\(\Omega\) [\si{\cubic\angstrom}/atom] & 24.39 & 24.09 & 25.81 & 25.33-25.95\tnote{c} \\
\bottomrule
\end{tabular*}
\begin{tablenotes}
    \item[a] \footnotesize \cite{gnut61zfaac,sang97jpe, robb73jpap, baro89mcp} 
    \item[b] \footnotesize \cite{somm66jap}
    \item[c] \footnotesize \cite{leon03im,capu16jem}
\end{tablenotes}
\end{threeparttable}
\end{table}

In Table~\ref{tbl:1}, we report the optimized lattice parameters and the unit-cell volumes obtained for the binary crystals using PBE, SCAN, and HSE06 functionals.
These results are compared to experimental findings determined by x-ray and electron diffraction techniques~\cite{jack57prslsmps,sang97jpe,robb73jpap, baro89mcp,somm66jap,leon03im,capu16jem}. PBE leads to an overestimation of the lattice constant of Cs$_3$Sb and K$_3$Sb by about \SI{0.17}{\angstrom} and \SI{0.1}{\angstrom}, respectively. This is a common shortcoming of this approximation that has been extensively reported for various bulk materials \cite{zhan18njp,schi11jcp, tran16jcp}. In the case of Na$_3$Sb, we obtain a different trend, with the PBE result underestimating the experimental reference by about \SI{0.1}{\angstrom}.
Adopting HSE06 improves the accuracy for Cs$_3$Sb and in particular for K$_3$Sb, where the experimental reference is matched (see Table~\ref{tbl:1}).
For Na$_3$Sb, however, HSE06 leads to a further underestimation of the lattice parameter compared to PBE. 
It is worth noting that the FCC phase of this material manifests itself at high pressures in contrast to the hexagonal one that is stable at room conditions~\cite{leon03im,capu16jem}. We chose this structure rather than the hexagonal one for a direct comparison with the other binary crystals.
Lattice parameters and volumes of Cs$_3$Sb and Na$_3$Sb obtained with SCAN are within the experimental range. In the case of K$_3$Sb, the results obtained with SCAN are a few tenth m\si{\angstrom} below the experimental value but the improvement over PBE and the reduced costs compared to HSE06 make it the best choice.  

\begin{table}[h!]
\centering
\begin{threeparttable}
    \caption{Conventional lattice parameters ($a$ and $c$ for the hexagonal crystal only) and unit-cell volumes per atom (\(\Omega\))
    obtained with the SCAN functional for Cs\(_2\)KSb, CsK\(_2\)Sb, NaK\(_2\)Sb, and Na\(_2\)KSb. Available experimental values taken from Refs.~\cite{mcca65jpcs,mcca60jopacos} are reported in parentheses.}
    \label{tbl:2}
    \begin{tabular*}{0.9\textwidth}{@{\extracolsep{\fill}} lcccc}
    \hline
        & Cs\(_2\)KSb &  CsK\(_2\)Sb &  NaK\(_2\)Sb &  Na\(_2\)KSb \\
        \midrule
        \( a \) [\si{\angstrom}] & 9.11 & 8.62 (8.61\tnote{a}) & 5.58  (5.61\tnote{b}) & 7.60 (7.72\tnote{b}) \\
        \( c \) [\si{\angstrom}] &   &   & 10.92 (10.93\tnote{b})&   \\ 
        \( \Omega\) [\si{\angstrom^3}/atom] & 50.81 & 40.03 (39.89\tnote{a}) & 36.74 (37.31\tnote{b})& 27.44 (28.76\tnote{b})\\
        \\ \hline
        \bottomrule
    \end{tabular*}
    \begin{tablenotes}
        \item[a] \footnotesize \cite{mcca65jpcs}.
        \item[b] \footnotesize \cite{mcca60jopacos}.
    \end{tablenotes}
\end{threeparttable}
\end{table}

Having identified SCAN as the optimal xc functional to accurately describe the lattice parameters of the binary systems, we employ it to determine the structural properties (lattice constants and volumes) of the considered ternary alkali antimonides (see Table~\ref{tbl:2}). 
Excellent agreement is found for CsK$_\mathrm{2}$Sb where the calculated values match almost perfectly the experimental references. 
The results obtained for the in-plane ($a$) and out-of-plane ($c$) lattice parameters of NaK$_\mathrm{2}$Sb are in equally good agreement with measurements.
A slight discrepancy of 0.12~\AA{} with respect to the measurement is found for the lattice constant of Na$_\mathrm{2}$KSb (see Table~\ref{tbl:2}): however, it cannot considered detrimental due to the scarcity of experimental references. 
For Cs$_2$KSb, no direct measurements of the lattice constant are available. An interpolated value of 8.88~\AA{} is reported in Ref.~\cite{ette02prb}: although it is not far from our prediction (see Table~\ref{tbl:2}), we do not consider it a reliable point of comparison. 
The structural parameters obtained with SCAN for the ternary alkali antimonides considered in this work are in better agreement with the available experimental references than earlier DFT studies adopting semi-local functionals~\cite{wu23jmca,ette02prb,amad21jpcm,cocc19sr,sassn21es,schi22prm,shar23aaem,zhon21ijer}.

\subsection{Phonon dispersions}

\begin{figure*}
\centering
\includegraphics[width=1\textwidth]{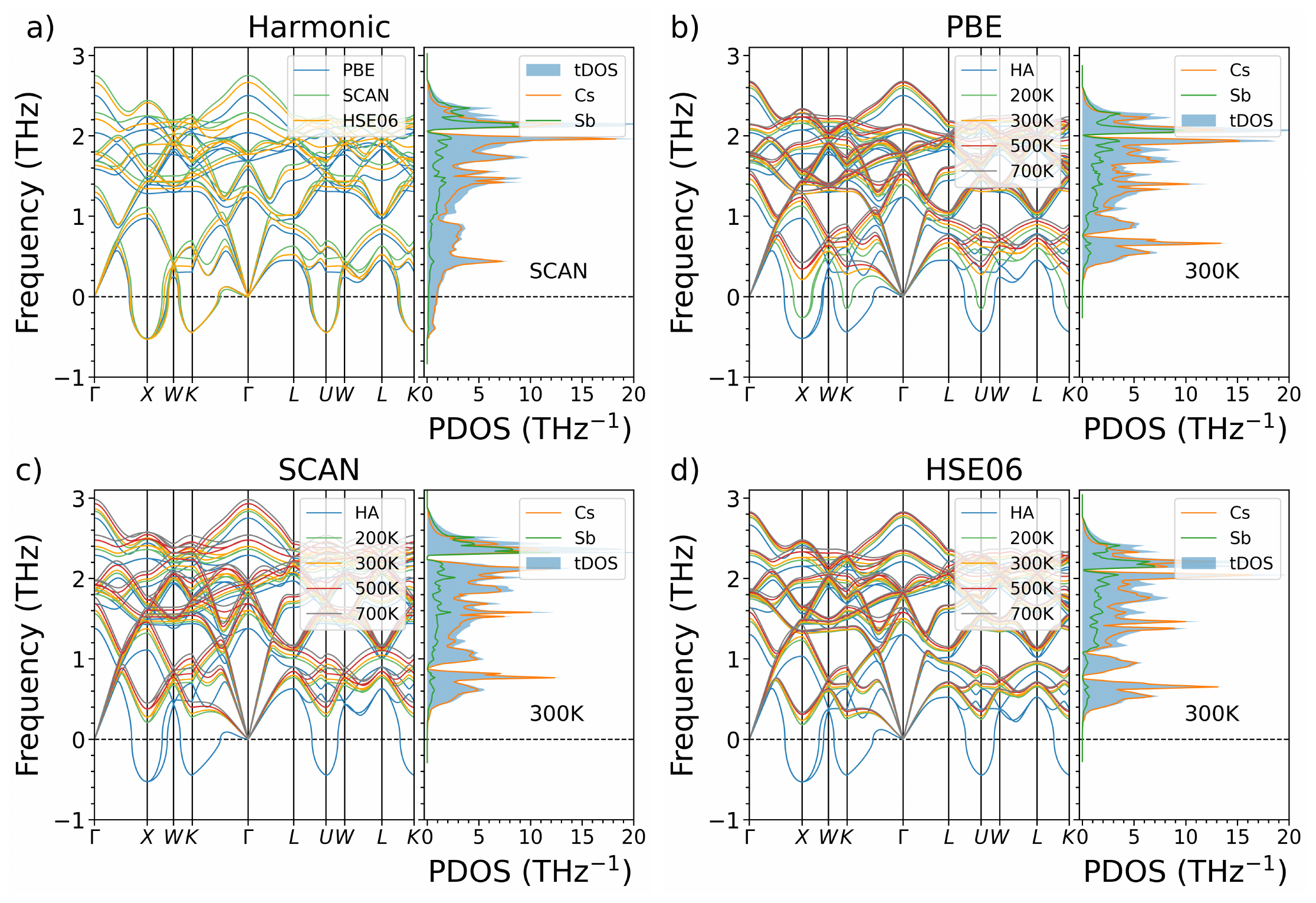}
\caption{a) Harmonic phonon band structure (left) and atom-projected projected density of states (PDOS) of Cs$_\mathrm{3}$Sb computed with different xc functionals.
Temperature-dependent phonon dispersion of Cs$_\mathrm{3}$Sb obtained with b) PBE, c) SCAN, and d) HSE06 in the harmonic (HA) dispersion and with anharmonic contributions at 200~K, 300~K, 500~K, and 700~K. On the right side of panels b)-d), the anharmonic phonon PDOS at 300~K is reported.}
\label{fig:panel_Cs3Sb}
\end{figure*}

To assess the performance of the three considered xc functionals in the calculation of the vibrational properties of multi-alkali antimonides, we evaluate the phonon dispersions of the binary systems. The harmonic approximation predicts a dynamic instability of Cs$_\mathrm{3}$Sb revealed by imaginary frequencies (displayed as negative values) ranging from 0 to 0.7i~THz with minima at the high-symmetry points X, K, and U (see Figure~\ref{fig:panel_Cs3Sb}a). This behavior is independent of the choice of the functional, although the imaginary minima obtained with SCAN are higher in frequency compared to those delivered by PBE and HSE06.
By inspecting the harmonic PDOS on the right side of Figure~\ref{fig:panel_Cs3Sb}a, we identify the predominant contribution of Cs atoms in the low-frequency region, while Sb atoms have a stronger influence at higher frequencies. This behavior is consistent with the masses of the two atomic species.

Phonon frequencies are closely related to the bond strengths: the tighter the interactions, the higher the frequencies. The known over-softening featured by PBE results can be attributed to two main factors: (i) the underestimation of bond strengths as a consequence of overestimated lattice parameters (see Table~\ref{tbl:1}) and (ii) the tendency of this functional to overestimate the electronic polarizability (see Table 1 in the Supporting Information), which amplifies the resonant bonding characteristics~\cite{lee14nc}.  
Both SCAN and HSE06 correct the over-softening of PBE (see Figure \ref{fig:panel_Cs3Sb}a), due to their more accurate assessment of intermediate- and long-range interactions, as demonstrated in previous studies~\cite{zhan17prb,sun16nc,shao23prb}. 

Using SCPH, we calculate the anharmonic phonon dispersions of Cs$_\mathrm{3}$Sb at increasing temperatures using the three different functionals, see Figures~\ref{fig:panel_Cs3Sb}b-d. Corrections from quartic, quintic, and sextic IFCs are included and, regardless of the chosen form for $v_{xc}$, they lift the dynamical instability featured in the harmonic approximation~\cite{zhon21ijer}.
However, the temperature at which imaginary phonon frequencies disappear from the spectra depends on the adopted functional.
With PBE, the phonon dispersion computed at 200~K is qualitatively analogous to the one obtained with harmonic contributions only (Figure~\ref{fig:panel_Cs3Sb}b), and the dynamical instability is lifted at 230~K.
With SCAN and HSE06, this threshold temperature is below 200~K and equal to 170~K and 190~K, respectively (see Figures~\ref{fig:panel_Cs3Sb}c,d).
The most significant differences in the phonon dispersions calculated with different functionals appear at the $\Gamma$ point and are mainly due to longitudinal/transverse (LO/TO) splitting of the optical phonons. These changes reflect alterations in electronic polarization, dielectric tensor, and Born effective charges, as expected from the different functionals~\cite{skel15jcp}. 
The results obtained with SCAN and HSE06 for Cs$_\mathrm{3}$Sb are overall very similar both in the harmonic approximation as well as when anharmonic corrections are included (Figure~\ref{fig:panel_Cs3Sb}c-d). 
Nonetheless, due to its lower computational costs, comparable to those of semi-local DFT, the SCAN functional represents the optimal choice to perform these calculations.

\begin{figure*}
\centering
\includegraphics[width=1\textwidth]{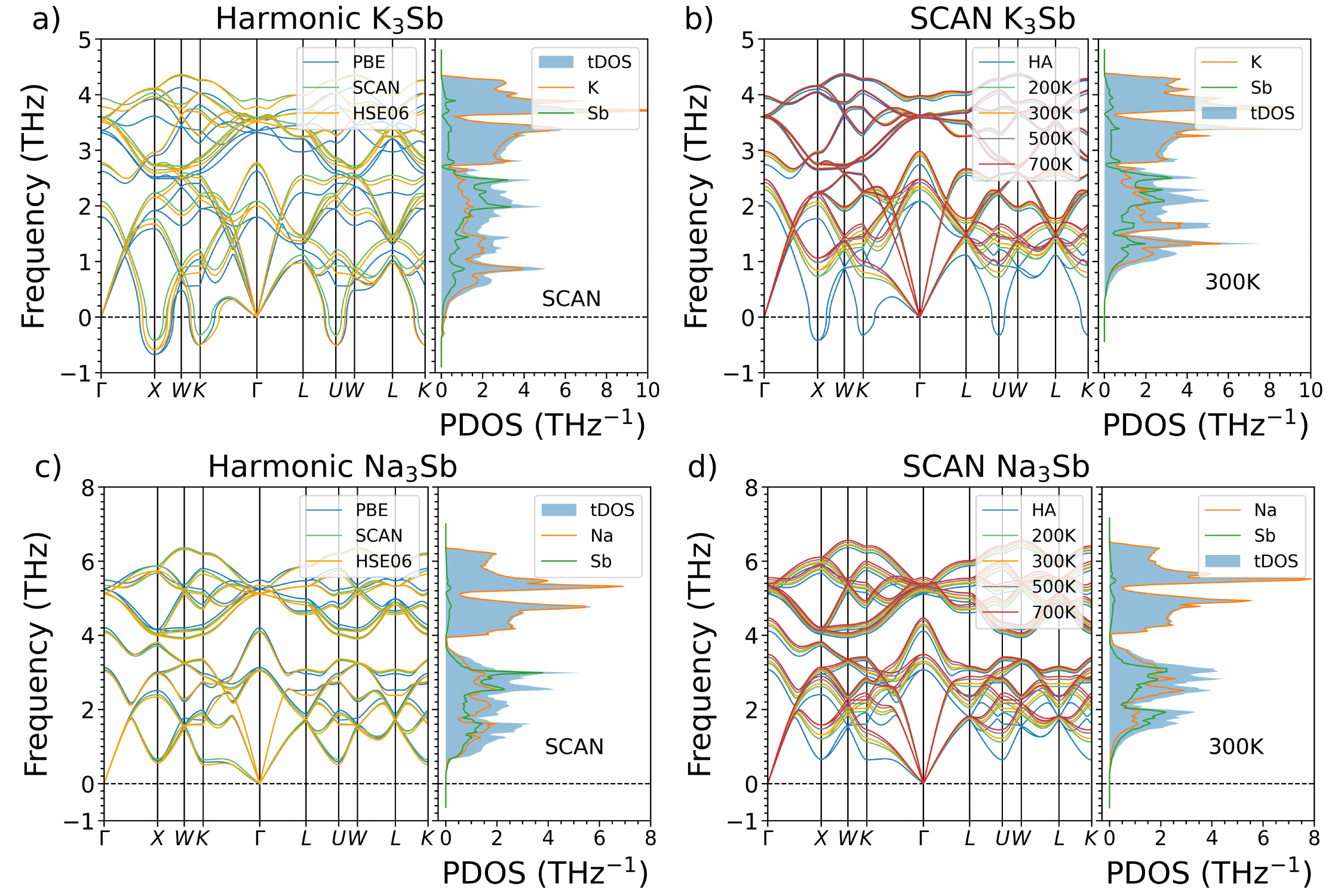}
\caption{Harmonic phonon dispersion computed with different xc functionals for a) K$_\mathrm{3}$Sb and c) Na$_\mathrm{3}$Sb, and temperature-dependent phonon dispersion (result in the harmonic approximation, HA, shown for comparison) obtained with SCAN for b) K$_\mathrm{3}$Sb and d) Na$_\mathrm{3}$Sb.}
\label{fig:panel_K3Sb_Na3Sb}
\end{figure*}

The phonon dispersion of K$_\mathrm{3}$Sb in the harmonic approximation (see Figure~\ref{fig:panel_K3Sb_Na3Sb}a), closely resembles that of Cs$_\mathrm{3}$Sb (compare Figure~\ref{fig:panel_Cs3Sb}a), with imaginary frequencies emerging within the same range and exhibiting minima at identical high-symmetry points. The comparison among results obtained with different functionals highlights differences in the LO/TO splitting at the $\Gamma$ point. For instance, by inspecting the PDOS, it is evident that the lighter atom K contributes more significantly at higher frequencies (see Figure~\ref{fig:panel_K3Sb_Na3Sb}a, right panel) compared to Cs (Figure~\ref{fig:panel_Cs3Sb}a), as expected.
The temperature-dependent anharmonic phonon dispersion of K$_\mathrm{3}$Sb obtained with SCAN (see Figure~\ref{fig:panel_K3Sb_Na3Sb}b) reveals that the inclusion of higher-order IFCs leads to the disappearance of imaginary frequencies at a lower temperature compared to Cs$_\mathrm{3}$Sb. Temperature-dependent phonon dispersions for K$_\mathrm{3}$Sb obtained with HSE06 and PBE are similar to those obtained with SCAN (see Figure~S1 in the Supporting Information).

For Na$_\mathrm{3}$Sb, the situation is different.
In the harmonic approximation (Figure~\ref{fig:panel_K3Sb_Na3Sb}c), the absence of imaginary frequencies highlights the dynamic stability of this material in contrast to all the other binary alkali antimonides considered in this work and in the literature~\cite{zhao20prb}. This characteristic of Na$_3$Sb can be attributed to the small atomic radius of Na, which reduces steric hindrance and repulsive interactions, contributing to a more stable structural arrangement. Furthermore, examining the PDOS in Figure~\ref{fig:panel_K3Sb_Na3Sb}c, it is evident that Na contributes more than Sb at higher frequencies, in contrast to heavier alkali atoms that contribute especially in the low-frequency range.
Na$_\mathrm{3}$Sb shows a weaker temperature dependence compared to other binary compounds, as shown in Figure~\ref{fig:panel_K3Sb_Na3Sb}d. Due to the more pronounced harmonic characteristics of the material, intermediate and long-range interactions are less significant compared to heavier alkali antimonides, leading to minor disparities among results obtained with different xc functional (see Figure~S2 in the Supporting Information). We also checked that the qualitative features of the phonon PDOS of the binary alkali antimonides considered in this work are rather insensitive to temperature (see Figure~S3).

\begin{figure*}[h!]
\centering
\includegraphics[width=1\textwidth]{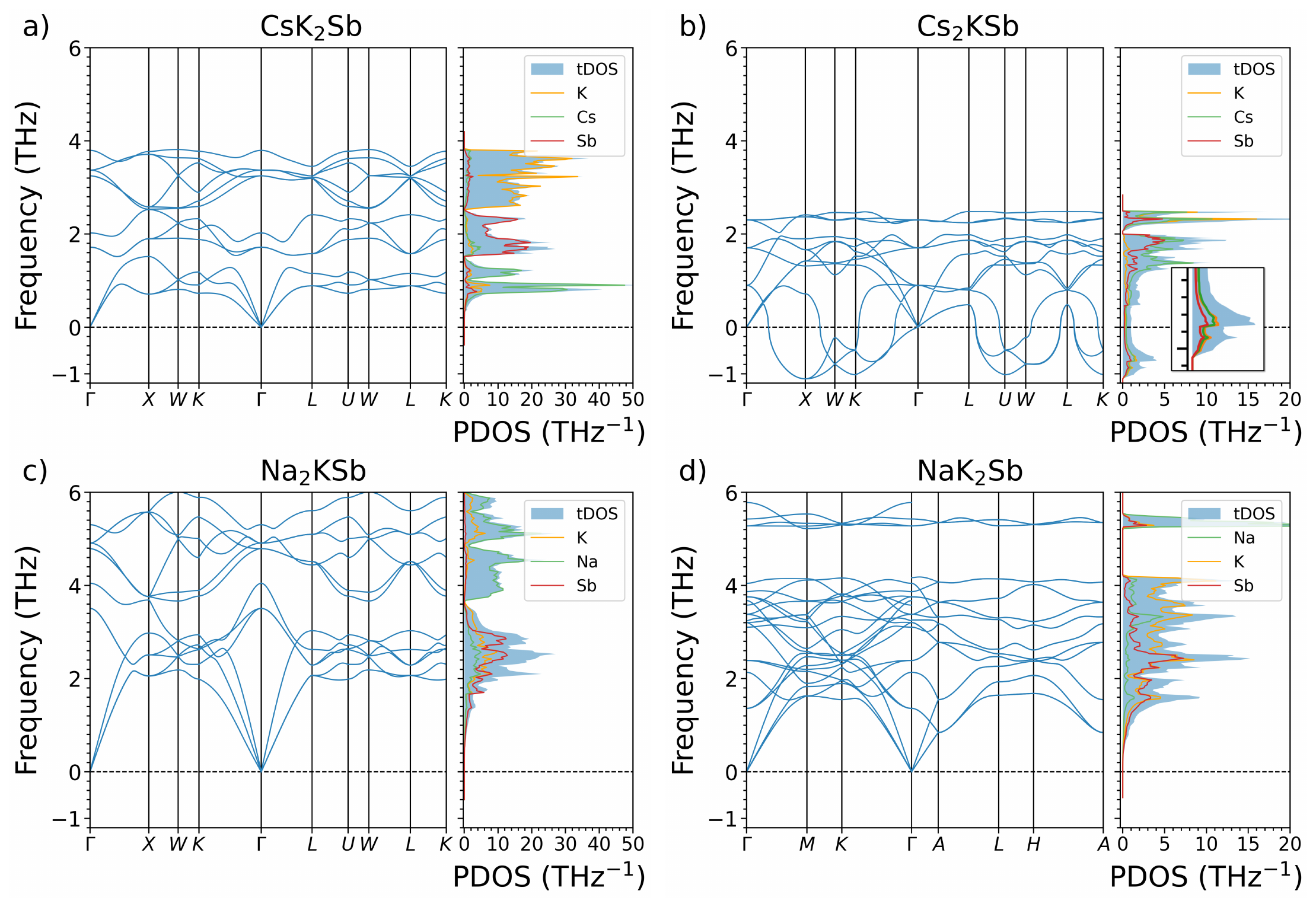}
\caption{Harmonic phonon dispersions and PDOS of a) CsK$_\mathrm{2}$Sb, b) Cs$_\mathrm{2}$KSb, c) Na$_\mathrm{2}$KSb, and d) NaK$_\mathrm{2}$Sb computed at 0~K with the SCAN functional.}
\label{fig:panel_mat}
\end{figure*}

With the insight gained from the binary systems, we now proceed with the analysis of the ternary compounds adopting only the SCAN functional, which offers the optimal trade-off between accuracy and computational efficiency.  
In Figure~\ref{fig:panel_mat}, we show the phonon band structures and PDOS computed at 0~K in the harmonic approximation. Except for Cs$_\mathrm{2}$KSb, which exhibits imaginary phonon frequencies (Figure~\ref{fig:panel_mat}b), the other three crystals are dynamically stable. In the following, we discuss their vibrational features in detail.
The phonon spectrum of CsK$_\mathrm{2}$Sb (Figure \ref{fig:panel_mat}a) is characterized by three distinct regions. The acoustic modes forming the lowest-energy band (0 - 1.4~THz) are dominated by Cs contributions, while the low- and high-frequency optical modes (1.4 - 2.2~THz and 2.2 - 3.6~THz) are mainly related to Sb and K atoms, respectively, in line with their atomic masses. A clear separation between acoustic and optical branches occurs along the $\Gamma$–X, $\Gamma$–K, and $\Gamma$–L directions. This characteristic, termed ``avoided crossing'', flattens the corresponding phonon bands and reduces the group velocity of the acoustic phonons and, consequently, the lattice thermal conductivity, thereby enhancing the electronic and thermal transport properties of the material~\cite{shar23aaem,yuan22jmcc}. In CsK$_\mathrm{2}$Sb, a distinct separation between optical and acoustic modes is observed, with an evident LO/TO splitting at the $\Gamma$ point. This feature suggests strong long-range electrostatic interactions within its lattice.
The occurrence of imaginary modes within the harmonic approximation for CsK$_\mathrm{2}$Sb is documented in previous DFT studies~\cite{yuan22jmcc}, where the stabilization of this system was achieved by incorporating higher-order force constants using the PBEsol functional. In contrast to these results, our study adopting the SCAN functional does not reveal any imaginary frequencies in the harmonic approximation, highlighting the importance of a suitable xc potential to obtain reliable results.

The phonon dispersion for Cs$_2$KSb (see Figure \ref{fig:panel_mat}b), exhibits imaginary frequencies ranging from 0 to 1.1i THz, with notable minima at the high-symmetry points X, K, U, and W. This finding points to a dynamical instability of Cs$_2$KSb mainly influenced by Cs and K atoms, as evidenced in the PDOS on the right side of Figure~\ref{fig:panel_mat}b. 
Although Cs$_2$KSb shares the same crystal structure as CsK$_2$Sb, the instability of the former material is enhanced by the predominance of Cs atoms over K. Notably, Cs atoms exhibit a larger atomic displacement parameter than K~\cite{yuan22jmcc}. This larger displacement from the equilibrium position of the Cs atoms leads to a ``rattling", characterized by large amplitude vibrations of weakly bound atoms. This behavior contributes to the softening of the acoustic phonon branches and introduces pronounced anharmonic effects, as has been already reported in other DFT studies~\cite{yuan22jmcc, maug18jmcc}.

Na$_\mathrm{2}$KSb is structurally identical to its Cs-based counterpart but in contrast to the latter, it is dynamically stable. 
The phonon dispersion of this material, shown in Figure~\ref{fig:panel_mat}c, features two main regions.
Both acoustic and optical modes are found in the low-frequency range (0 - 3.5~THz).
The latter are dominated by Sb modes with non-negligible contributions from the alkali species, while all atoms participate almost equally in the acoustic modes.
The optical modes at higher frequencies (3.5 - 5.8~THz) are mainly due to Na vibrations with K contributions becoming more relevant above 5~THz; the effect of Sb atoms is negligible in this region.
The phonon dispersion of Na$_2$KSb exhibits a significant LO/TO splitting, which is the most prominent among the ternary compounds investigated in this work. This feature hints at strong long-range electrostatic forces within the lattice structure of Na$_2$KSb, in agreement with previous DFT studies~\cite{yue22prb}. This prominence in LO/TO splitting is also indicative of a heightened ionic character in Na$_2$KSb, which aligns with the high static dielectric constant (see Table S1). 

In Figure~\ref{fig:panel_mat}d, the DFT phonon dispersion of hexagonal NaK$_\mathrm{2}$Sb is reported for the first time, according to the best of our knowledge.
Due to the larger number of atoms in this crystal compared to the cubic ones analyzed above, many more modes appear in the band structure.
Acoustic and optical modes are seamlessly found in the region 0 - 4.5~THz with mostly K-Sb and Na-Sb contributions below 3~THz and between 3.5 and 4.5~THz, and with a K predominance around 3~THz. 
A separate manifold of optical phonon bands, mostly related to Na-centered modes, appears around 5~THz. 
Overall, this hexagonal crystal exhibits flatter phonon bands compared to cubic Na$_2$Kb. However, the atomic contributions in the various frequency regions are consistent in the two materials, confirming that this characteristic is mainly related to the composition rather than to the stoichiometry and/or the crystalline arrangement.

\begin{figure*}[h!]
\centering
\includegraphics[width=0.67\textwidth]{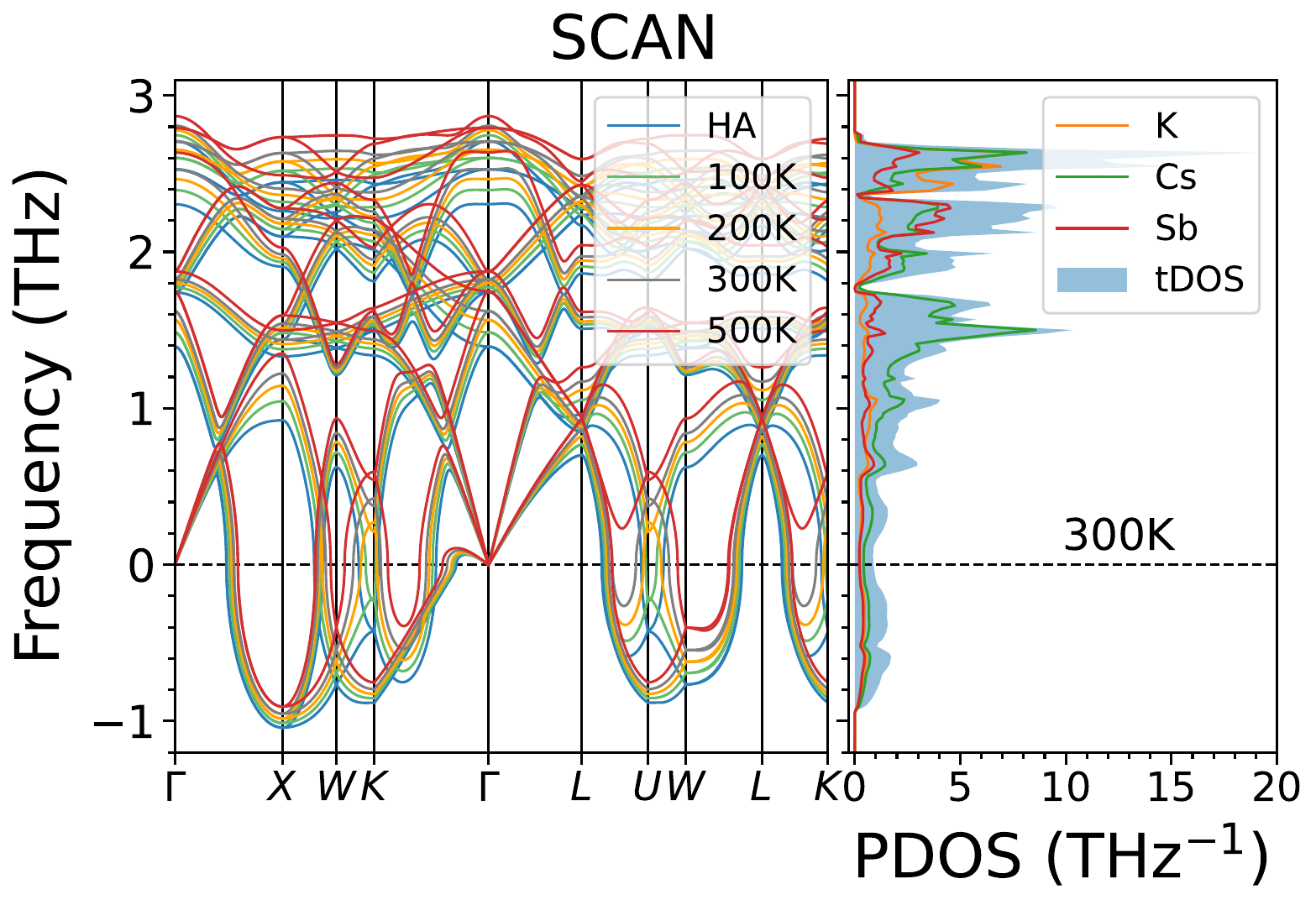}
\caption{Phonon dispersion of Cs$_\mathrm{2}$KSb obtained with the SCAN functional in the harmonic approximation (HA) and including anharmonic contributions at 100~K, 200~K, 300~K, and 500~K. The atom-resolved PDOS displayed on the right panel corresponds to the anharmonic result at 300~K.}
\label{fig:Cs2KSb_anharmonic}
\end{figure*}

In the last part of this study, we deepen the analysis on Cs$_\mathrm{2}$KSb, which is dynamically unstable at 0~K (see Figure~\ref{fig:panel_mat}b).
To this end, we computed the anharmonic phonon dispersion of this material at increasing temperatures including cubic, quartic, quintic, and sextic IFCs. All results shown in Figure~\ref{fig:Cs2KSb_anharmonic} feature imaginary modes confirming the instability of Cs$_\mathrm{2}$KSb regardless of temperature.
A trend of increasing (negative) frequencies is seen with increasing temperature but even at 500~K, the lowest minima remain well below zero (0.8i~THz).
The persistence of imaginary modes indicates the inherent dynamic instability of the crystal, which occurs because the structure is located at a local maximum or a saddle point in the potential energy surface (PES). To explore other possible stable configurations, it is crucial to conduct a thorough examination of the PES, for example, by perturbing the crystal structure along a specific phonon eigenvector. This procedure, commonly known as ``mode-mapping", enables sampling the PES based on the amplitude of the harmonic mode~\cite{skel16prl} and will be the subject of dedicated future work. 
These findings obtained for Cs$_\mathrm{2}$KSb suggest that, in contrast to the Na-based counterpart which is predicted to be stable already at 0~K, this material is unstable even at and beyond room temperature and thus not favored during growth. 
This outcome is in agreement with empirical notions~\cite{schm18prab}.

\section{Conclusions}
In summary, we investigated the thermodynamic and vibrational properties of binary and ternary alkali antimonide crystals by means of DFT and machine learning methods going beyond the harmonic approximation. We assessed the performance of different exchange-correlation functionals, including PBE, HSE06, and SCAN, in determining the structural parameters and the phonon dispersion of the binary systems. Having found that SCAN offers the optimal trade-off between accuracy and computational efforts,  we used it to study the ternary compounds. 
The phonon band structures and the atom-resolved density of states reveal a clear correlation between Cs content and dynamical stability. Specifically, crystals with a higher Cs concentration are less stable than the others and exhibit imaginary phonon frequencies. While in Cs$_\mathrm{3}$Sb these instabilities are resolved going beyond the harmonic approximation, Cs$_\mathrm{2}$KSb remains unstable even above room temperature, matching the observation that this compound does not form upon evaporation. More in-depth studies exploring the potential energy surface of this compound are required to identify a stable structure. In general, we found that an increase in the atomic mass of the alkali metals leads to a decrease in the energies of the low-lying phonon modes associated with these species. We also provided the first DFT results for the phonon properties of hexagonal NaK$_\mathrm{2}$Sb, which appears to be stable already in the harmonic approximation.

To conclude, our results offer valuable indications about the thermodynamic stability of multi-alkali antimonides. The insight gained from the analysis of their phonon dispersions and density of states can contribute to refining the growth recipes of these materials as well as the prediction of their photoemission yield. From a theoretical perspective, the benchmark of the DFT functional on the thermodynamic and vibrational properties of alkali antimonides complements the findings of an earlier study focused only on their electronic characteristics~\cite{sassn21es} and promotes SCAN as the optimal choice for the study of these materials even in a high-throughput framework~\cite{cocc21m,sass22jcp}.

\section*{Acknowledgement}
This work was funded by the German Research Foundation, project numbers 490940284 and 182087777 (CRC 951), by the German Federal Ministry of Education and Research (Professorinnenprogramm III), and by the State of Lower Saxony (Professorinnen f\"ur Niedersachsen, SMART, and DyNano). J.S.A. acknowledges support from the Evonik Stiftung. The computational resources were provided by the North-German Supercomputing Alliance (HLRN), project nic00076, and by the high-performance computing cluster CARL at the University of Oldenburg, funded by the German Research Foundation (Project No. INST 184/157-1 FUGG) and by the Ministry of Science and Culture of the State of Lower Saxony. 


\section*{References} 
\bibliographystyle{unsrt}

\end{document}